\documentstyle[preprint,epsf,pre,aps]{revtex}
\newcommand{\grad}{\nabla}
\newcommand{\be}{\begin{eqnarray}}
\newcommand{\ee}{\end{eqnarray}}

\begin{document}
\date{\today}
\draft

\title{The von Karman equations, the stress function,
and elastic ridges in high dimensions.}
\author{Eric M. Kramer\cite{byline}}
\address
{The James Franck Institute and the Department of Physics\\
The University of Chicago\\
Chicago, Illinois 60637}
\maketitle

\begin{abstract}

The elastic energy functional 
of a thin elastic rod or sheet is generalized
to the case of an $M$-dimensional manifold 
in $N$-dimensional space. We derive potentials
for the stress field and curvatures and find the 
generalized von Karman equations for a manifold
in elastic equilibrium. 
We perform a scaling analysis of 
an $M-1$ dimensional ridge in an $M=N-1$ dimensional
manifold. A ridge of linear size $X$ in a 
manifold with thickness $h \ll X$ has a width 
$w \sim h^{1/3}X^{2/3}$ and a total energy 
$E \sim \mu h^{M} (X/h)^{M-5/3}$, where $\mu$ is
a stretching modulus.
We also prove that the total bending
energy of the ridge is exactly five times
the total stretching energy. 
These results match those of A. Lobkovsky 
[Phys. Rev. E {\bf 53}, 3750 (1996)] for the 
case of a bent plate in three dimensions.
 
\end{abstract}


\section{Introduction}

The crumpling of a thin elastic sheet
is mediated by the formation of a network of 
narrow ridges \cite{Kram3}. Plastic deformation 
of the material in the neighborhood of these
ridges leads to the ubiquitous linear scars 
in crushed paper, aluminum foil, and car bodies 
\cite{Crmpl,Car,Sound}.
It was recently discovered by Witten, Lobkovsky,
and others that this phenomenon can be 
accounted for using linear elasticity theory \cite{WitLi,Lob1,Lob2}.
The scaling laws for a ridge
were first derived by Witten and Li using a Flory type argument.
A ridge of length $X$ in a sheet of thickness
$h$ and Young's modulus $Y$ was found
to have a total elastic energy 
$E \sim Y h^{3}(X/h)^{1/3}$ and 
a width $w \sim h^{1/3} X^{2/3}$.
Lobkovsky confirmed this result with a scaling analysis
of the von Karman equations describing a thin,
semi-infinite strip with a single ridge.
He also verified
these results with detailed simulations.

One important result of this analysis is the
discovery that the stresses
and curvatures decay rapidly to zero  
in the direction transverse to the ridge.
The length scale of this decay is the ridge width.
We therefore interpret ridge formation as a
{\it confinement} of the elastic stress field \cite{Corr}.
Although there is a qualitative appreciation that
confinement is the result of the competition between
the in-plane strains and the curvatures of the plate,
a deeper theoretical understanding is still lacking. 
In particular there is no proof
of confinement under generic boundary conditions.
We have been working towards this goal. In a companion 
paper we prove that a thin plate must 
have regions of nonzero strain if it is to fit
into a small sphere \cite{Kram1}. It remains to be shown that
these regions necessarily assemble themselves into
a network of ridges.

To understand the causes and consequences of stress
confinement, it is useful to examine the higher
dimensional analogs of a crumpled sheet. 
For example, a thin plate crumpled in four
dimensions doesn't have to stretch, and we expect
there is no stress confinement in this case.
In this paper we present the simplest
field theory describing the strains and curvatures 
of a deformed elastic manifold in higher dimensions.
We also perform a scaling analysis for a ridge 
in these systems.

There is considerable precedent for examining
higher dimensional systems for insights into
membrane elasticity. Most notable are
studies of the so-called ``crumpling transition''
of a thin elastic membrane in thermal equilibrium
\cite{MemInt,CT1,CT2}. That work focuses on the way thermal 
fluctuations and self-avoidance renormalize
the elastic constants of a thin elastic
sheet. The field theory is often developed
in perturbation theory around a convenient,
higher dimension. In this paper we work
exclusively at zero temperature and the word
``crumpling'' refers to compression by external forces.

In elasticity theory it is common to approximate
a thin plate by its center surface, or {\it centroid}
\cite{Love,LL}.
A three-dimensional plate is thereby described 
using a two-dimensional manifold.
The elastic energy functional for the centroid is
found by integrating out the components of the stress
and strain tensors which are transverse to the 
long directions.
The mathematical analysis of these
approximations and their range of validity is
the subject of the theory of thin elastic shells
\cite{Shell1,Shell2,Shell3}. In this paper we use the standard methods 
of shell theory to derive the elastic
energy of an $M$-dimensional manifold embedded in
$N$-dimensional space. We treat the manifold as 
the centroid of an $N$-dimensional elastic solid 
with an infinitesimal thickness $h$ in $N-M$ directions. 
The resulting energy functional 
has pieces quadratic in the strains, curvatures,
and torsions 
of the manifold. We take a functional derivative 
of the energy to find the equations of static equilibrium.
For a plate in three dimensions these equations are
called the von Karman equations, first written down by
Theodore von Karman in 1910 \cite{vK}. We will refer 
to our general result by the same name. 
Our derivation has several new elements.

It is noteworthy that the 
elastic energy of a thin plate may be
written in terms of two scalar potentials. 
The {\it stress function} $\chi$, introduced by Airy in 1863,
is the source of in-plane stresses \cite{Airy,Seung}.
The {\it bending potential} $f$ is the source of curvatures.
Similarly, studies of a deformed solid 
often use the tensor stress function
$\chi_{\alpha\beta}$, introduced by Maxwell in 1870
\cite{Maxwell,Kleinert}. 
In this paper we present the generalization
of these potentials for arbitrary $M$ and $N$.
Our derivation of the von Karman
equations reveals a role for the stress function
as the Lagrange multiplier of a geometric 
constraint in the energy functional.
For $M>2$ the stress function is a gauge field.

Next we turn our attention to the ridge structure. 
The width of the ridge in a plate scales
like $w \sim h^{1/3}X^{2/3}$, so in the limit $h \ll X$
the ridge is approximately one-dimensional and straight.
Analogously, the ridges in an $M$-dimensional manifold
are expected to be approximately $(M-1)$-dimensional
and to have no curvature. Indeed, simulations of a solid
elastic ball crushed by a sphere in four dimensions show
that the elastic energy is concentrated 
into flat, planar structures \cite{Kram3}.
As mentioned above, Lobkovsky has done a thorough analysis of an
isolated ridge for the case of a semi-infinite
sheet in three dimensions \cite{Lob1}. 
We repeat his analysis 
for the case of an $M$-dimensional, semi-infinite manifold
bent into a ridge in $M+1$-dimensional space.
We find that a ridge with linear size
$X$ has a width $w \sim h^{1/3}X^{2/3}$ 
and a total energy $E \sim Y h^{N} (X/h)^{M-5/3}$. Lastly, 
we prove that the total energy due to the curvature of the ridge is 
exactly five times the energy due to the strains.
These results match correctly onto the solution
for a bent plate in three dimensions \cite{Lob1,LobWit}. 

In Sec. II we review the 
differential geometry of a weakly strained 
$M$-dimensional manifold embedded in $N$-dimensional space.
In Sec. III we derive the elastic energy
of this manifold as the thin limit of an
$N$-dimensional elastic solid.
In Sec. IV we present the generalization of 
the stress function and the bending potential.
Then we make a variational
derivation of the von Karman equations.
In Sec. V we generalize the scaling
analysis of Lobkovsky to a ridge in $M >2$.
In Sec. VI we summarize our conclusions.

\section{Differential geometry review}

In this section we review the differential geometry 
of a weakly strained $M$-dimensional manifold ${\cal M}$ 
embedded in $N$-dimensional space $\Re^{N}$. By weak strains 
we mean (1) the strains are small compared to unity
and (2) the derivatives of the strains are small 
compared to the other relevant inverse lengths
(curvatures and torsions). These are the usual assumptions
of thin plate theory \cite{LL}. Note that they do not
prohibit arbitrarily large deformations of the manifold. 
With these assumptions
the math simplifies considerably. A treatment
of the topics in this section using
the full apparatus of differential
geometry may be found in Refs. \cite{Diff1,Diff2}.
 
The manifold is flat in the 
absence of external forces, so it can be 
parameterized by the Euclidean coordinate patch
$\{ \vec{x}=x_{i}\hat{e}_{i} \in {\cal M}$ 
for $i \in [1,M]$ ; $x_{i}=0$ for $i \in [M+1,N] \}$
where $\{ \hat{e}_{i} \}$ are the Euclidean basis vectors.
We will refer to $(x_{\alpha} ; \alpha \in [1,M])$
as the {\it manifold} coordinate patch and 
denote it with Greek subscripts.
Any deformation of ${\cal M}$
can then be represented as a continuous 
map $\vec{r}(x_{\alpha})$ from the manifold coordinates to 
$\Re^{N}$. At each point on the deformed manifold
there is an $M$-dimensional tangent 
space spanned by the tangent vectors  
$\vec{t}_{\alpha}=\partial_{\alpha}\vec{r}$.
The metric on the manifold is then
$g_{\alpha\beta}= \vec{t}_{\alpha}\cdot\vec{t}_{\beta}$
and the strain tensor is
$u_{\alpha\beta}=(1/2)(g_{\alpha\beta}-\delta_{\alpha\beta})$.

We treat all relevant quantities to lowest order
in the strains.
This immediately gives us 
$g_{\alpha\beta}=\delta_{\alpha\beta}+O(u)$,
so there is no need to distinguish
the covariant components of a tensor 
from the contravariant components.
The Christoffel symbols are
$\Gamma_{\alpha\beta}^{\gamma}=\partial_{\beta}u_{\alpha\gamma}
+\partial_{\alpha}u_{\beta\gamma}
-\partial_{\gamma}u_{\alpha\beta}+O(u^{2})$ \cite{Diff2}. 
Thus, $D_{\alpha}=\partial_{\alpha}+O(u)$
and covariant derivatives are just partial derivatives
to leading order. 
Geodesics are approximately straight lines in the 
manifold coordinates.

The extrinsic curvature tensor for the manifold is defined
$\vec{K}_{\alpha\beta}=D_{\alpha}\vec{t}_{\beta}
\approx \partial_{\alpha}\vec{t}_{\beta}$.
It is straightforward to show that the components
of this tensor are normal to the tangent space. We start with 
\be
\vec{t}_{\alpha}\cdot\vec{K}_{\beta\gamma} & = &
\vec{t}_{\alpha}\cdot\partial_{\beta}
\vec{t}_{\gamma} \nonumber \\
& = & \partial_{\beta}(\vec{t}_{\alpha}
\cdot\vec{t}_{\gamma})-
\partial_{\beta}\vec{t}_{\alpha}\cdot
\vec{t}_{\gamma} \nonumber \\
& = & \partial_{\beta}(\delta_{\alpha\gamma})
-\vec{K}_{\alpha\beta}\cdot\vec{t}_{\gamma} 
\nonumber \\
& = & -\vec{t}_{\gamma}\cdot\vec{K}_{\alpha\beta}
\ee
This quantity is therefore odd under a cyclic
permutation of the indices. Three such permutations
gives
$\vec{t}_{\alpha}\cdot\vec{K}_{\beta\gamma}
=-\vec{t}_{\alpha}\cdot\vec{K}_{\beta\gamma}=0$.

We choose a set of orthonormal basis vectors 
$\{\hat{n}^{(\alpha)}(x_{\beta})\}$ to span
the ($N-M$)-dimensional normal space
at each point on the manifold. Note that
we use Greek-in-parenthesis for the
normal index $(\alpha)\in [M+1,N]$. In this basis
the extrinsic curvature tensor becomes 
$\vec{K}_{\alpha\beta}=C_{\alpha\beta}^{(\gamma)}\,\hat{n}^{(\gamma)}$
where $C_{\alpha\beta}^{(\gamma)}=
\hat{n}^{(\gamma)}\cdot\vec{K}_{\alpha\beta}$ and
summation over repeated indices is implied.
We will refer to the $N - M$ tensors 
$C_{\alpha\beta}^{(\gamma)}$ as the normal 
components of the extrinsic curvature tensor.
Note that $C_{\alpha\beta}^{(\gamma)}=C_{\beta\alpha}^{(\gamma)}$
and $\vec{K}_{\alpha\beta}=\vec{K}_{\beta\alpha}$ 
since $\vec{K}_{\alpha\beta}
\approx \partial_{\alpha}\partial_{\beta}\vec{r}$. 

It is useful to expand the derivatives
of the normal vectors in the full basis 
$\{ \vec{t}_{\alpha},\hat{n}_{(\alpha)} \}$ 
\be
\partial_{\alpha}\hat{n}^{(\beta)}
=-C_{\alpha\gamma}^{(\beta)}\,\vec{t}_{\gamma}
-\tau^{(\beta)(\gamma)}_{\alpha}\,\hat{n}^{(\gamma)}
\label{eq:Wein}
\ee
where we have defined the torsions 
$\tau^{(\beta)(\gamma)}_{\alpha}(x)=
-\hat{n}^{(\gamma)}\cdot\partial_{\alpha}\hat{n}^{(\beta)}$
and where we have used
$0=\partial_{\alpha}(\vec{t}_{\gamma}\cdot\hat{n}^{(\beta)})
=\vec{t}_{\gamma}\cdot\partial_{\alpha}\hat{n}^{(\beta)}
+C_{\alpha\gamma}^{(\beta)}$.
Eq. (\ref{eq:Wein}) is the generalization of the
Weingarten map for a plate \cite{Diff2}.
Taking one derivative of 
$\hat{n}^{(\alpha)}\cdot \hat{n}^{(\beta)}=\delta_{(\alpha)(\beta)}$
gives us the antisymmetry property for the torsions
$\tau^{(\beta)(\gamma)}_{\alpha}=-\tau^{(\gamma)(\beta)}_{\alpha}$.

The last quantity we will need is the 
intrinsic curvature tensor 
$R_{\alpha\beta\mu\nu}=\vec{K}_{\alpha\mu}\cdot\vec{K}_{\beta\nu}
-\vec{K}_{\alpha\nu}\cdot\vec{K}_{\beta\mu}$ \cite{Diff2}.
In the normal basis, this becomes
\be
R_{\alpha\beta\mu\nu}[C]=
C_{\alpha\mu}^{(\gamma)}C_{\beta\nu}^{(\gamma)}
-C_{\alpha\nu}^{(\gamma)}C_{\beta\mu}^{(\gamma)}
\label{eq:R}
\ee
The intrinsic curvature tensor is related to the strain
tensor via the generalization of Gauss'
{\it Theorema Egregium}
\be
R_{\alpha\beta\mu\nu}[u]=-u_{\alpha\mu,\beta\nu}
+u_{\alpha\nu,\beta\mu}-u_{\beta\nu,\alpha\mu}
+u_{\beta\mu,\alpha\nu}+O(u^{2})
\label{eq:TE}
\ee
where indices to the right of a comma
indicate partial derivatives.
Eqs. (\ref{eq:R}) and (\ref{eq:TE}) together
give $R_{\alpha\beta\mu\nu}[C]=R_{\alpha\beta\mu\nu}[u]$.
This is one version of the {\it geometric}
von Karman equation, so called because 
it expresses the geometric constraint relating
the extrinsic curvature and the strain.
It is straightforward to verify this equation by
substituting the definition of the strain tensor
into $R_{\alpha\beta\mu\nu}[u]$ and differentiating.

The intrinsic curvature tensor will be most useful
to us in the linear combination 
\be
G_{\alpha\beta}=R_{\alpha\nu\beta\nu}
-\frac{1}{2}\delta_{\alpha\beta}R_{\mu\nu\mu\nu}
\ee 
This is the Einstein curvature tensor, familiar from
general relativity \cite{GR}. It is symmetric 
$G_{\alpha\beta}=G_{\beta\alpha}$ and
satisfies the conservation law
$\partial_{\alpha}G_{\alpha\beta}=0$.
Taking the appropriate contractions of the 
geometric von Karman equation gives us
\be
G_{\alpha\beta}[C]=G_{\alpha\beta}[u]
\label{eq:gvK1}
\ee

\section{The elastic energy functional}

In this section we obtain an expression
for the elastic energy
of an $M$-dimensional manifold ${\cal M}$ via the thin limit
of an $N$-dimensional solid ${\cal N}$.
Versions of this calculation for a rod and plate may be found in 
Refs. \cite{Love} and \cite{LL}.

To begin we consider the elastic energy 
functional for an arbitrary $N$-dimensional solid.
We keep the assumption of small strains used in the last
section, but we relax the condition on the 
derivatives. As before, there is a 
Euclidean coordinate patch $(x_{i})$ covering the 
undeformed manifold 
$\{ \vec{x}=x_{i}\hat{e}_{i} \in {\cal N}$ 
for $i \in [1,N] \}$. This is the {\it material}
coordinate patch, denoted by Latin indices. 
Under the application of external forces the solid
assumes an embedding $\vec{r}(x)$.
The tangent space of the solid
is the full $\Re^{N}$ and the tangent vectors
are $\vec{t}_{i}(x)=\partial_{i}\vec{r}$. The metric
is $g_{ij}(x)=\vec{t}_{i}\cdot\vec{t}_{j}$ and the strain
tensor is $u_{ij}=(1/2)(g_{ij}-\delta_{ij})$. 
 
Two consequences of $u_{ij} \ll 1$ are 
(1) the volume element of the deformed solid
$dr^{N}(x)$ is well-approximated by the volume element
of the undeformed solid $dx^{N}$ and 
(2) the elastic energy of the material 
only needs to be calculated to second 
order in the strains. The most general energy functional
quadratic in the strains and consistent with an isotropic 
material is 
\be
E[u]=\int_{{\cal N}} dx^{N} \left( \mu u_{ij}^{2} 
+ \frac{\lambda}{2} u_{ii}^{2} \right)
\label{eq:Eij}
\ee
where $\mu$ and $\lambda$ are the Lam\'{e} coefficients \cite{LL}.
It is useful to rewrite this equation
$E=(1/2)\int dx^{N} \sigma_{ij} u_{ij}$
where 
\be
\sigma_{ij}(x)= 2\mu u_{ij} +\lambda \delta_{ij} u_{kk}
\label{eq:sigu}
\ee
is the stress field conjugate to $u_{ij}$.
The stress field satisfies 
the conservation law $D_{i}\sigma_{ij}=0$.
 
By analogy with the treatments of a rod and sheet, 
we assume that 
${\cal N}={\cal M}\times B^{N-M}(h)$
where $B^{N-M}(h)$ is an ($N - M$)-ball of 
infinitesimal radius $h$.
The choice of a spherical ``cross-section''
is important to preserve the full rotational
symmetry in the normal space of ${\cal M}$.  
The material coordinate patch becomes
$\{ \vec{x}=x_{i}\hat{e}_{i}\in {\cal M}$ 
for $i \in [1,M]$ ; $\sum_{i=M+1}^{N} x_{i}^{2} \leq h^{2} \}$.
We refer to the long directions as 
the {\it manifold} coordinates and denote
them with Greek indices. We refer to
the short directions as the {\it normal} or
{\it transverse} coordinates and denote 
them with Greek-in-parenthesis. 
For clarity we relabel the 
transverse coordinates $\zeta_{(\alpha)}=x_{(\alpha)}$,
so that $(x_{i})=(x_{1},x_{2},\ldots x_{M},
\zeta_{(M+1)},\zeta_{(M+2)},\ldots \zeta_{(N)})$.
 
The M-dimensional surface satisfying $\zeta_{(\alpha)}=0$
is the {\it centroid} of ${\cal N}$.
When the transverse degrees of freedom are integrated out 
it is the centroid which becomes the manifold ${\cal M}$.
Under the application of external forces the 
centroid deforms to some equilibrium
embedding $\vec{r}^{c}(x)=\vec{r}(x,\zeta=0)$.
All the quantities discussed in the previous section are 
well-defined with respect to this embedding.
With the exception of $C^{(\lambda)}_{\alpha\beta}$ and
$\tau^{(\sigma)(\lambda)}_{\alpha}$, 
we will denote quantities calculated on the centroid 
with a superscript $c$.

The first step in deriving the 
elastic energy functional for the centroid is to 
make a Taylor expansion of the
embedding $\vec{r}(x,\zeta)$ in $\zeta$  
\be
\vec{r}(x,\zeta)=\vec{r}^{c}(x)
+\zeta_{(\mu)}\vec{a}^{(\mu)}(x)
+\frac{1}{2}\zeta_{(\mu)}\zeta_{(\nu)}
\,\vec{b}^{(\mu)(\nu)}(x)+\cdots
\ee
where 
$\vec{a}^{(\mu)}(x)=\partial_{(\mu)}
\vec{r}(x,\zeta)|_{\zeta=0}$ and
$\vec{b}^{(\mu)(\nu)}(x)=\partial_{(\mu)}\partial_{(\nu)}
\vec{r}(x,\zeta)|_{\zeta=0}$.
Recall that in Sec. II we had the freedom
to choose an arbitrary set of torsions
due to the rotational symmetry of the normal space. 
Here we make the natural assignment 
$\hat{n}^{(\mu)}(x) = \vec{a}^{(\mu)}/|\vec{a}^{(\mu)}|$.
With this identification the torsions of the normal basis
are the torsions of the deformed solid.

To calculate the energy 
we need to solve for $\vec{a}^{(\mu)}$ and $\vec{b}^{(\mu)(\nu)}$
in terms of $u^{c}_{\alpha\beta}$, $C^{(\mu)}_{\alpha\beta}$,
$\tau^{(\mu)(\nu)}_{\alpha}$, and their derivatives.
We make the following assumptions: (1)
$u^{c}_{\alpha\beta} \ll 1$,
$C^{(\lambda)}_{\alpha\beta} \ll 1/h$, and
$\tau^{(\sigma)(\lambda)}_{\alpha} \ll 1/h$.
We will see that these are necessary to satisfy the
small strain condition $u_{ij} \ll 1$.
(2) The smallest length scale $\ell$ over which
the strains, curvatures, and torsions vary satisfies $\ell \gg h$.
We therefore write the most general 
expressions consistent with the rotational
and reflection symmetries of the problem
to first nontrivial order
\be
\vec{a}^{(\mu)} & = & (1+a_{1} u^{c}_{\alpha\alpha})\,\hat{n}^{(\mu)}
\nonumber \\
\vec{b}^{(\mu)(\nu)} & = & b_{1} (C^{(\mu)}_{\alpha\alpha}
\,\hat{n}^{(\nu)} + C^{(\nu)}_{\alpha\alpha}\,\hat{n}^{(\mu)})
+ b_{2} \delta^{(\mu)(\nu)} (C^{(\lambda)}_{\alpha\alpha}
\,\hat{n}^{(\lambda)}) 
\label{eq:ab}
\ee
where $a_{1}$, $b_{1}$, and $b_{2}$ are dimensionless constants
to be determined. The corrections
are of relative order $O(u, hC, h\tau, \mbox{and} h/\ell)$.
A more detailed account of the derivation 
of Eq. (\ref{eq:ab}) may be found in Appendix A.
Note that $\vec{a}^{(\mu)}$ and $\vec{b}^{(\mu)(\nu)}$ 
are independent of $\vec{t}_{\alpha}$
and $\tau^{(\mu)(\nu)}_{\alpha}$, primarily because 
the torsions are antisymmetric under $\mu\leftrightarrow\nu$.
If the cross-section of ${\cal N}$ isn't rotationally
symmetric, then there are additional possibilities in 
Eq. (\ref{eq:ab}). This may couple the torsions
to the curvatures in a nontrivial way and
complicate the resulting theory considerably. 

The tangent vectors to first nontrivial order are 
\be
\partial_{\alpha}\vec{r} & = & \vec{t}^{c}_{\alpha}+
\zeta_{(\mu)}\{-C_{\alpha\gamma}^{(\mu)}\,\vec{t}_{\gamma}
-\tau_{\alpha}^{(\mu)(\gamma)}\,\hat{n}^{(\gamma)} \}
\nonumber \\
\partial_{(\alpha)}\vec{r} & = & 
(1+a_{1} u^{c}_{\gamma\gamma})\,\hat{n}^{(\alpha)}
+ b_{1} \zeta_{(\mu)} (C^{(\alpha)}_{\alpha\alpha}
\,\hat{n}^{(\mu)} + C^{(\mu)}_{\gamma\gamma}\,\hat{n}^{(\alpha)})
+b_{2} \zeta_{(\alpha)}(C^{(\lambda)}_{\gamma\gamma}
\,\hat{n}^{(\lambda)})
\ee
and the components of the strain tensor are
\be
\begin{array}{lll}
u_{\alpha\beta} & = & u^{c}_{\alpha\beta}-\zeta_{(\mu)}
C_{\alpha\beta}^{(\mu)} 
\\
u_{\alpha (\beta)} & = & -\frac{1}{2}\zeta_{(\mu)}
\tau_{\alpha}^{(\mu)(\beta)}
 \\
u_{(\alpha)(\beta)} & = & a_{1} u^{c}_{\gamma\gamma} 
\delta_{(\alpha)(\beta)}
+b_{1} \zeta_{(\mu)} (\delta_{(\alpha)(\beta)}
C_{\gamma\gamma}^{(\mu)})
+ \displaystyle\frac{b_{1}+b_{2}}{2} 
(\zeta_{(\alpha)}C_{\gamma\gamma}^{(\beta)}
+ \zeta_{(\beta)}C_{\gamma\gamma}^{(\alpha)})
\end{array}
\label{eq:u1}
\ee
From this it is clear that our assumption (1) above  
is equivalent to the small strain condition $u_{ij} \ll 1$.
Also note that the transverse derivatives
of the strain tensor $\partial_{(\alpha)}u_{ij}$ 
are of $O(C, \tau)$ and are not negligible. Only the
manifold derivatives $\partial_{\alpha}u_{ij}$ 
can be safely neglected to leading order, as
assumed in Sec. II.

We derive the energy for a portion of the manifold
far from the regions where external forces 
are applied. We therefore have the boundary condition
$\sigma_{(\alpha)(\beta)}|_{\zeta=h}=0$. Combined with 
the conservation law $D_{i}\sigma_{ij}=0$,
we have $\sigma_{(\alpha)(\beta)}=0$ everywhere.
This condition specifies $u_{(\alpha)(\beta)}$ uniquely.
Referring to Eq. (\ref{eq:sigu})  
\be
\sigma_{(\alpha)(\beta)}= 2 \mu u_{(\alpha)(\beta)}
+\lambda\delta_{(\alpha)(\beta)} u_{kk} =0 
\label{eq:tmp1}
\ee
so $u_{(\alpha)(\beta)} \sim \delta_{(\alpha)(\beta)}$ and
$b_{1}+b_{2}=0$.
Substituting the trace of the strain tensor 
\be
u_{kk}=u^{c}_{\alpha\alpha} - \zeta_{(\mu)}C_{\alpha\alpha}^{(\mu)}
+(N-M)(a_{1}u^{c}_{\alpha\alpha} + b_{1}\zeta_{(\mu)}
C_{\alpha\alpha}^{(\mu)})   
\ee
into Eq. (\ref{eq:tmp1}) gives us  
$a_{1}=-b_{1}=- c_{0}$
where $c_{0} = \lambda/(2\mu +(N-M)\lambda)$.
Thus
\be
u_{(\alpha)(\beta)} & = & -c_{0}\delta_{(\alpha)(\beta)}
(u^{c}_{\gamma\gamma}-\zeta_{(\mu)}C_{\gamma\gamma}^{(\mu)})
\label{eq:u2} \\
u_{kk} & = & c_{1}(u^{c}_{\gamma\gamma}-\zeta_{(\mu)}
C_{\gamma\gamma}^{(\mu)})
\label{eq:u3}
\ee
where $c_{1}=2\mu/(2\mu +(N-M)\lambda)$.

Substituting Eqs. (\ref{eq:u1}) and (\ref{eq:u3})
into the expression for the strain Eq. (\ref{eq:sigu}) gives
\be
\begin{array}{lll}
\sigma_{\alpha \beta} & = & 2\mu (u^{c}_{\alpha\beta}-\zeta_{(\mu)}
C_{\alpha\beta}^{(\mu)}) +c_{1}\lambda\delta_{\alpha\beta}
(u^{c}_{\gamma\gamma}-\zeta_{(\mu)}C_{\gamma\gamma}^{(\mu)})
\\
\sigma_{\alpha (\beta)} & = & -\mu \zeta_{(\mu)}
\tau_{\alpha}^{(\mu)(\beta)}
\\
\sigma_{(\alpha)(\beta)} & = & 0
\end{array}
\ee
The elastic energy Eq. (\ref{eq:Eij}) becomes
\be
\begin{array}{rl}
\displaystyle E = \int_{{\cal M}} dx^{M} \int_{B^{N-M}(h)} 
d\zeta^{N-M} & \left\{ \mu (u^{c}_{\alpha\beta}-\zeta_{(\mu)}
C_{\alpha\beta}^{(\mu)})^{2} 
\right. \\
& \displaystyle \left. 
+c_{1}\frac{\lambda}{2}
(u^{c}_{\gamma\gamma}-\zeta_{(\mu)}C_{\gamma\gamma}^{(\mu)})^{2}
+\frac{\mu}{4}(\zeta_{(\mu)}\tau_{\alpha}^{(\mu)(\beta)})^{2}
\right\}
\end{array}
\ee
Since the transverse coordinates are being integrated 
over $B^{N-M}$, terms odd in $\zeta_{(\mu)}$ vanish and
we have finally
\be
\begin{array}{rl}
E[u,C,\tau] = \displaystyle \int_{{\cal M}} dx^{M} & \left\{ 
\mu^{c} \left( (u^{c}_{\alpha\beta})^{2} 
+ c_{0} (u^{c}_{\alpha\alpha})^{2}\right) \right. \\
& \displaystyle \left. 
+ \kappa \left( C_{\alpha\beta}^{(\mu)}C_{\alpha\beta}^{(\mu)}
+ c_{0} C_{\alpha\alpha}^{(\mu)}C_{\beta\beta}^{(\mu)} \right)
+ \frac{\kappa}{4} (\tau_{\alpha}^{(\mu)(\nu)}
\tau_{\alpha}^{(\mu)(\nu)})
\right\}
\end{array}
\ee
where  
\be
\mu^{c} & = & \mu \int_{B^{N-M}(h)} d\zeta^{N-M}
\label{eq:muc}
\ee
is the effective stretching modulus and
\be
\kappa & = & \mu \int_{B^{N-M}(h)} d\zeta^{N-M} \zeta_{(1)}^{2}
\label{eq:kappac}
\ee
is the effective bending modulus of the thin manifold ${\cal M}$.
The integral in Eq. (\ref{eq:muc}) is just the volume
of a sphere with radius $h$
\be
\int_{B^{d}(h)} d\zeta^{N-M} =\frac{1}{d}h^{d} S_{d}
\ee
where $S_{d} = 2\pi^{d/2}/\Gamma(d/2)$ is the area of a unit
sphere in d dimensions.
The integral in Eq. (\ref{eq:kappac}) is 
\be
\int_{B^{d}(h)} d\zeta^{d} \zeta_{(1)}^{2}
=\left\{ \begin{array}{lc}
\frac{2}{3}h^{3}    & d=1 \\
\frac{\pi}{4}h^{4} & d=2 \\
\frac{1}{d+2} h^{d+2} B(3/2,d-2) S_{d-1} & d >2
\end{array} \right.
\ee
where $B(a,b)= \Gamma(a)\Gamma(b)/\Gamma(a+b)$ is the beta
function \cite{Stegun}. 

We can rewrite the elastic energy using conjugate fields
\be
E[u,C,\tau] = \frac{1}{2} \int_{{\cal M}} dx^{M} \left\{ 
\sigma^{c}_{\alpha\beta}u_{\alpha\beta}
+ M^{(\mu)}_{\alpha\beta}C^{(\mu)}_{\alpha\beta}
+ T_{\alpha}^{(\mu)(\nu)}\tau_{\alpha}^{(\mu)(\nu)}
\right\}
\label{eq:Etot}
\ee
where 
\be
\sigma^{c}_{\alpha\beta}(x)=2\mu^{c} (u^{c}_{\alpha\beta}
+c_{0}\delta_{\alpha\beta}u^{c}_{\gamma\gamma})
\label{eq:sigu2}
\ee
is the resultant strain field,
\be
M^{(\mu)}_{\alpha\beta}(x)=2\kappa  (C^{(\mu)}_{\alpha\beta}
+ c_{0}\delta_{\alpha\beta}C^{(\mu)}_{\gamma\gamma}) 
\ee
is the bending moment field, and 
\be
T_{\alpha}^{(\mu)(\nu)}(x)=\frac{1}{2}\kappa\tau_{\alpha}^{(\mu)(\nu)} 
\ee
is the torsional moment field.
Eq. (\ref{eq:Etot}) is the full elastic energy 
functional for a thin elastic manifold.
We will frequently refer to the term quadratic
in the strains as the stretching energy
and the term quadratic in the curvatures
as the bending energy.

We henceforth drop the superscript $c$ 
and assume that all quantities refer to the
centroid manifold ${\cal M}$.

\section{The potentials and the von Karman equations}

\subsection{The case $M>2$}

Note that Eq. (\ref{eq:Etot}) does not explicitly
couple the strains to the curvatures of the manifold.
The strains and the curvatures are implicitly 
coupled because they are both defined via derivatives of the
embedding $\vec{r}(x)$. However, a naive functional derivative 
of Eq. (\ref{eq:Etot}) with respect to $u_{\alpha\beta}$
gives the trivial and incorrect result $\sigma_{\alpha\beta}=0$.
Previous authors, working with a thin plate
in three dimensions, have solved this problem by working in a special
coordinate system which is approximately tangent to
one point on the manifold. In this frame, known as the Monge
representation, the embedding is 
$\vec{r}(x)=[x_{\alpha}+u_{\alpha}(x),
w(x)]$ and the derivatives of $\vec{u}(x)$ and
$w(x)$ are assumed to be small everywhere.
To leading nontrivial order the strain tensor is 
$u_{\alpha\beta}=(1/2)(u_{\alpha,\beta}+u_{\beta,\alpha}
+w_{,\alpha}w_{,\beta})$ and the extrinsic curvature tensor
is $C_{\alpha\beta}=w_{,\alpha\beta}$.
Functional derivatives are then
taken with respect to $u_{\alpha}$ and $w$
and the correct equations are obtained \cite{Monge}.

We choose to work instead with the 
field variables $u_{\alpha\beta}$, $C^{(\mu)}_{\alpha\beta}$,
and their potentials. The 
advantages are (1) we work exclusively in the manifold
coordinates, so there is no need for an approximately
tangent frame, (2) it is easier to treat the 
boundary conditions, and (3) we discover a new
interpretation for the stress functions of Airy and 
Maxwell \cite{Airy,Maxwell,Kleinert}.
Due to some small differences, 
we focus here on the case $M>2$ and 
return to the case $M=2$ in the next subsection.
The coupling between the strain and the curvature
is completely accounted for by the geometric von
Karman equation Eq. (\ref{eq:gvK1}). We therefore add the 
Lagrange multiplier term 
\be
E_{\chi}[u,C,\chi] = \frac{1}{2} \int_{{\cal M}} dx^{M} 
\,\chi_{\alpha\beta}(G_{\alpha\beta}[C]-G_{\alpha\beta}[u])
\label{eq:Echi}
\ee
to the total elastic energy Eq. (\ref{eq:Etot}).
We will see that the Lagrange multiplier 
$\chi_{\alpha\beta}(x)$ is the tensor stress function.

One may ask why it is sufficient to use the Einstein
curvature tensor $G_{\alpha\beta}$ 
rather than the full intrinsic
curvature tensor $R_{\alpha\beta\mu\nu}$.
$G_{\alpha\beta}$ is symmetric in $\alpha\leftrightarrow\beta$ 
and constrained by the conservation law
$G_{\alpha\beta,\alpha}=0$, so a naive count of
the independent degrees of freedom gives $M(M-1)/2$.
$R_{\alpha\beta\mu\nu}$
is symmetric in $\alpha\leftrightarrow\beta$ 
and $\mu\leftrightarrow\nu$, and antisymmetric
in $(\alpha\beta)\leftrightarrow(\mu\nu)$, so in principle
it has $M(M-1)(M^{2}-M+2)/8$ independent components.
However, Eq. (\ref{eq:TE}) shows that for small strains 
the intrinsic curvature tensor is linear in 
the strain tensor $u_{\alpha\beta}$.
The strain tensor is symmetric and constrained
by the conservation of the resultant stress tensor,
so it has $M(M-1)/2$ independent components.
The Einstein curvature tensor is therefore the most
economical choice. The alternative 
forms for the Lagrange multiplier term
$\chi_{\alpha\beta\mu\nu}R_{\alpha\beta\mu\nu}$
and $\chi_{\alpha\beta}R_{\alpha\mu\beta\mu}$
both yield the correct von Karman equations
for the fields $\sigma_{\alpha\beta}$ and 
$C^{(\lambda)}_{\alpha\beta}$, but the Lagrange multiplier
is not identical to the stress function.

If the normal basis has zero torsion we can define 
a bending potential for each normal component of the 
extrinsic curvature tensor.
In the remainder of this paper we assume 
there are no external torsional moments
acting on the manifold. Because the torsions are not coupled
to the strains or to the curvatures in Eqs. (\ref{eq:Etot})
and (\ref{eq:Echi}), the solution is simply
$\tau_{\alpha}^{(\mu)(\nu)}(x)=0$.
With this we can prove the Codazzi-Mainardi relation
\be
\partial_{\alpha}C_{\beta\gamma}^{(\lambda)} 
& = & 
\hat{n}^{(\lambda)}\cdot (\partial_{\alpha}\vec{K}_{\beta\gamma})
+(\partial_{\alpha}\hat{n}^{(\lambda)})\cdot\vec{K}_{\beta\gamma}
\nonumber \\
& = &  \hat{n}^{(\lambda)}\cdot 
(\partial_{\alpha}\partial_{\beta}\vec{t}_{\gamma})
+(-C_{\alpha\delta}^{(\lambda)}\,\vec{t}_{\delta})
\cdot \vec{K}_{\beta\gamma}
\nonumber \\
& = & 
\hat{n}^{(\lambda)}\cdot (\partial_{\beta}\vec{K}_{\alpha\gamma})
\nonumber \\
&  = & \partial_{\beta}C_{\alpha\gamma}^{(\lambda)}
\label{eq:CM} 
\ee
We have used the simplified Weingarten map
$\partial_{\alpha}\hat{n}^{(\lambda)}=
-C_{\alpha\beta}^{(\lambda)}\,\vec{t}_{\beta}$
in the second line and the orthogonality condition 
$\vec{t}_{\alpha}\cdot\hat{n}^{(\beta)}=0$ in the third line.
The Codazzi-Mainardi relation is analogous to the zero-curl condition
on a vector field. It allows the definition of a
scalar potential
$f^{(\lambda)}(x)$ via $C_{\alpha\beta}^{(\lambda)}=\partial_{\alpha}
\partial_{\beta}f^{(\lambda)}$.

There is a novel
form for the Einstein curvature tensor which greatly 
simplifies the variational derivatives taken below.
We begin by defining the {\it double curl} operator,
valid for $M >2$,
\be
(d.c.)_{\alpha\beta\mu\nu} & = & \frac{1}{(M-3)!}\,
\epsilon_{\alpha\gamma\mu\tau_{1}\cdots\tau_{M-3}}
\epsilon_{\beta\delta\nu\tau_{1}\cdots\tau_{M-3}}
\partial_{\gamma}\partial_{\delta}
\nonumber \\
& = & \delta_{\alpha\beta}\delta_{\mu\nu}\grad^{2}
-\delta_{\alpha\nu}\delta_{\beta\mu}
\grad^{2} -\delta_{\alpha\beta}\partial_{\mu}\partial_{\nu}
\nonumber \\
& & +\delta_{\alpha\nu}\partial_{\beta}\partial_{\mu}
+\delta_{\beta\mu}\partial_{\alpha}\partial_{\nu}
-\delta_{\mu\nu}\partial_{\alpha}\partial_{\beta}
\ee
where $\epsilon_{\tau_{1}\cdots\tau_{M}}$ is the Levi-Civita tensor  
\be
\epsilon_{\tau_{1}\cdots\tau_{M}}=\left\{\begin{array}{ll}
+1 & (\tau_{1},\tau_{2},\ldots ,\tau_{M}) \mbox{  even permutation
of  } (1,2,3,\ldots,M) \\
-1 & (\tau_{1},\tau_{2},\ldots ,\tau_{M}) \mbox{  odd permutation
of  } (1,2,\ldots,M) \\
0  & \mbox{otherwise}
\end{array} \right.
\ee
The double curl is antisymmetric in $\alpha\leftrightarrow\mu$
and $\beta\leftrightarrow\nu$ and symmetric in 
$(\alpha\mu)\leftrightarrow (\beta\nu)$ (compare
to $R_{\alpha\beta\mu\nu}$). It satisfies 
$\partial_{\alpha}(d.c.)_{\alpha\beta\mu\nu}=0$ by construction.
The Einstein curvature tensor may be written 
$G[u]_{\alpha\beta} = (d.c.)_{\alpha\beta\mu\nu}u_{\mu\nu}$
or
$G[f]_{\alpha\beta} = (d.c.)_{\alpha\beta\mu\nu}
[(1/2)f_{,\mu}^{(\lambda)}f_{,\nu}^{(\lambda)}]$.
These expressions are easily verified by substitution.

Now we can write the full expression for the energy functional,
including the Lagrange multiplier 
\be
E[u,f,\chi] = \int_{{\cal M}} dx^{M} \frac{1}{2} & \left\{
\sigma_{\alpha\beta}[u]u_{\alpha\beta}
+ \kappa (f_{,\mu\nu}^{(\lambda)}f_{,\mu\nu}^{(\lambda)}
+c_{0}f_{,\mu\mu}^{(\lambda)}f_{,\nu\nu}^{(\lambda)})
\right\} \nonumber \\ &
+\chi_{\alpha\beta}(d.c.)_{\alpha\beta\mu\nu}
(\frac{1}{2}f_{,\mu}^{(\lambda)}f_{,\nu}^{(\lambda)}-u_{\mu\nu})
\label{eq:Epot}
\ee
Taking a functional variation with respect to $u_{\alpha\beta}$
and integrating the Lagrange multiplier term twice by parts gives
\be
\begin{array}{rl}
\delta E = \displaystyle\int_{{\cal M}} dx^{M} & 
\delta u_{\mu\nu} (\sigma_{\mu\nu}
-(d.c.)_{\mu\nu\alpha\beta}\chi_{\alpha\beta}) 
\\
& +\frac{1}{(M-3)!}
\epsilon_{\alpha\gamma\mu\tau_{1}\cdots\tau_{M-3}}
\epsilon_{\beta\delta\nu\tau_{1}\cdots\tau_{M-3}}
\partial_{\gamma}(\delta u_{\mu\nu} \,\chi_{\alpha\beta,\delta}
-\delta u_{\mu\nu,\delta} \,\chi_{\alpha\beta})
\end{array}
\label{eq:dEdu}
\ee
where we have used the symmetry of $u_{\alpha\beta}$ and
$\chi_{\alpha\beta}$.
The first term in Eq. (\ref{eq:dEdu}) gives a
conservation law for the resultant stress tensor.
Taking $\delta E/\delta u_{\alpha\beta}=0$, 
\be
\sigma_{\alpha\beta}=(d.c.)_{\alpha\beta\mu\nu}\,\chi_{\mu\nu}
\label{eq:sigchi}
\ee 
This is a restatement of the conservation
law $\partial_{\alpha}\sigma_{\alpha\beta}=0$.
Eq. (\ref{eq:sigchi}) is the defining
equation for Maxwell's stress function in $M=3$ 
\cite{Maxwell,Kleinert}.
We see that the stress function is a Lagrange
multiplier. This interpretation persists even
when $M=N=3$ and the extrinsic curvature tensor is identically zero
[take $f^{(\lambda)}=0$ in Eq. (\ref{eq:Epot})]. 
Eq. (\ref{eq:sigchi}) also provides a natural 
generalization of the stress function to higher
dimensions. 

One can verify by substitution that the stress tensor remains
unchanged under the local gauge transformations
$\chi_{\alpha\beta} \rightarrow \chi_{\alpha\beta}
+(1/2)(\xi_{\alpha,\beta}+\xi_{\beta,\alpha})$
where $\xi_{\alpha}(x)$ is an arbitrary vector field \cite{gauge}.
The tensor stress function is therefore a gauge field with
$M(M-1)/2$ physical degrees of freedom. This
agrees with the fact that the stress tensor
itself has $M(M-1)/2$ independent components. 

The second term in Eq. (\ref{eq:dEdu})
is a perfect differential. Using Gauss'
Law to rewrite it as an integral over the $M-1$
dimensional boundary of 
the manifold $\partial {\cal M}$ gives
\be
\delta E|_{\partial {\cal M}} = \frac{1}{(M-3)!}
\epsilon_{\alpha\gamma\mu\tau_{1}\cdots\tau_{M-3}}
\epsilon_{\beta\delta\nu\tau_{1}\cdots\tau_{M-3}}
\int_{\partial {\cal M}} dx^{M-1} \,\hat{m}_{\gamma}
(\delta u_{\mu\nu} \,\chi_{\alpha\beta,\delta}
-\delta u_{\mu\nu,\delta} \,\chi_{\alpha\beta})
\ee
where $\hat{m}$ is the unit outward normal
defined in the tangent space of ${\cal M}$.
The application of this term to a specific
problem depends on the boundary conditions
imposed at $\partial {\cal M}$.

Taking the functional variation of $E$ with respect to 
$f^{(\lambda)}$ and integrating by parts gives 
\be
\begin{array}{rl}
\delta E = \displaystyle\int_{{\cal M}} dx^{M} &
\delta \! f^{(\lambda)} \left( 2\kappa (1+c_{0}) \grad^{4}f^{(\lambda)} -
f^{(\lambda)}_{,\alpha\beta}\,(d.c.)_{\alpha\beta\mu\nu}
\chi_{\mu\nu} \right)
\\
& \displaystyle + 2 \kappa \,\partial_{\gamma} 
\left( \delta \! f^{(\lambda)}_{,\gamma} 
f^{(\lambda)}_{,\nu\nu}
+ c_{0}\, \delta \! f^{(\lambda)}_{,\nu} f^{(\lambda)}_{,\gamma\nu} 
-(1+c_{0})\, \delta \! f^{(\lambda)} f^{(\lambda)}_{,\gamma\gamma\nu} 
\right)
\\ 
& + \frac{1}{(M-3)!}\,
\epsilon_{\alpha\gamma\mu\tau_{1}\cdots\tau_{M-3}}
\epsilon_{\beta\delta\nu\tau_{1}\cdots\tau_{M-3}}
\,\partial_{\gamma} \left( \delta \! f^{(\lambda)} 
f^{(\lambda)}_{,\mu\nu} \,\chi_{\alpha\beta,\delta}
-\delta \! f^{(\lambda)}_{,\delta} f^{(\lambda)}_{,\mu\nu} 
\,\chi_{\alpha\beta} \right)
\end{array}
\ee
where there is no sum on $(\lambda)$.
The second and third terms are perfect differentials
and may be written as a condition on $\partial {\cal M}$.
Taking $\delta E/\delta f^{(\lambda)}=0$, the first term gives 
\be
2\kappa (1+c_{0}) \grad^{4}f^{(\lambda)}=
f^{(\lambda)}_{,\alpha\beta}\,(d.c.)_{\alpha\beta\mu\nu}
\,\chi_{\mu\nu}
\label{eq:fvK2}
\ee
which may be rewritten in the more familiar form
\be
M^{(\lambda)}_{\alpha\beta,\alpha\beta}=\sigma_{\alpha\beta}
\, C^{(\lambda)}_{\alpha\beta}
\label{eq:fvK1}
\ee
This is the {\it force} von-Karman equation, which expresses the 
balance of forces on a transverse section of the 
thin manifold.

To complete our discussion of the von Karman
equations, we write the Einstein curvature 
tensor in terms of the stress function.
First invert Eq. (\ref{eq:sigu2}) to get
\be
u_{\alpha\beta}=\frac{1}{2\mu}(\sigma_{\alpha\beta}
-c_{2}\delta_{\alpha\beta}\sigma_{\gamma\gamma})
\ee 
where $c_{2}=\lambda/(2\mu+N\lambda)$. Then 
\be
G_{\alpha\beta}[\chi]& =& \frac{1}{2\mu}\left\{
(d.c.)_{\alpha\beta\mu\nu}(d.c.)_{\mu\nu\sigma\tau}\chi_{\sigma\tau}
-c_{2}(d.c.)_{\alpha\beta\mu\mu}(d.c.)
_{\nu\nu\sigma\tau}\chi_{\sigma\tau} \right\}
\nonumber \\
& = & \frac{1}{2\mu}\left\{
(c_{3}-1)\left(\delta_{\alpha\beta}\grad^{4}\chi_{\sigma\sigma}
-\delta_{\alpha\beta}\grad^{2}\chi_{\sigma\tau,\sigma\tau}
-\grad^{2}\chi_{\sigma\sigma,\alpha\beta}\right)
\right. 
\nonumber \\
& & \left. +c_{3}\chi_{\sigma\tau,\alpha\beta\sigma\tau}
+\grad^{4}\chi_{\alpha\beta}-\grad^{2}\chi_{\alpha\sigma,\beta\sigma}
-\grad^{2}\chi_{\beta\sigma,\alpha\sigma}
\right\}
\ee 
where $c_{3}=(M-2)[2\mu +(N-M+2)\lambda]/(2\mu+N\lambda)$.

We thus have several alternative expressions for 
the von Karman equations, depending on which fields
are most convenient.
In terms of the extrinsic curvature and 
strain tensors, we have Eqs. (\ref{eq:gvK1}) and (\ref{eq:fvK1}).
In terms of the bending potentials and the stress 
function we have $G_{\alpha\beta}[f]=G_{\alpha\beta}[\chi]$ 
and Eq. (\ref{eq:fvK2}).

\subsection{The case $M =2$}

In this section we rederive the
von Karman equations for a thin plate.
Although the equations of the three-dimensional
problem have been discussed in detail
by several authors, our variational 
derivation of the force von Karman equation 
is particularly transparent \cite{Love,LL,Seung,Lob1}.
The only change from the previous 
section is that the tensor double curl operator
is not defined for $M =2$. Instead we use the scalar operator 
\be
(d.c.)_{\alpha\beta} =  \epsilon_{\alpha\mu}
\epsilon_{\beta\nu}\partial_{\mu}\partial_{\nu}
= \delta_{\alpha\beta}\grad^{2}-\partial_{\alpha}
\partial_{\beta}
\ee

When M $=2$, the intrinsic curvature 
tensor has only one independent component,
which we take as the generalization of the
Gaussian curvature 
\be
\kappa_{G}[C]=\frac{1}{2}R_{\alpha\beta\alpha\beta}[C]
=C^{(\lambda)}_{11}C^{(\lambda)}_{22}
-C^{(\lambda)}_{12}C^{(\lambda)}_{12}
\ee
It is straightforward to verify that 
$\kappa_{G}[u]=-(d.c.)_{\alpha\beta}u_{\alpha\beta}$ and 
$\kappa_{G}[f]=-(d.c.)_{\alpha\beta}\left(\frac{1}{2}
f^{(\lambda)}_{,\alpha}f^{(\lambda)}_{,\beta} \right)$. 

The elastic energy is
\be
E[u,C,\chi]  
=  \int_{{\cal M}} dx^{M} & \frac{1}{2}\left\{ 
\sigma_{\alpha\beta}[u]u_{\alpha\beta}
+ M^{(\lambda)}_{\alpha\beta}C^{(\lambda)}_{\alpha\beta} \right\}
+\chi (-\kappa_{G}[C]+\kappa_{G}[u])
\ee
\be
\begin{array}{rl}
\displaystyle E[u,f,\chi]  =  
\int_{{\cal M}} dx^{M} & 
\displaystyle \frac{1}{2} \left\{ 
\sigma_{\alpha\beta}[u]u_{\alpha\beta}
+ \kappa (f_{,\alpha\beta}^{(\lambda)}f_{,\alpha\beta}^{(\lambda)}
+c_{0}f_{,\beta\beta}^{(\lambda)}f_{,\beta\beta}^{(\lambda)})
\right\}
\\
& \displaystyle   +\chi (d.c.)_{\alpha\beta}
\left( \frac{1}{2}f_{,\alpha}^{(\lambda)}f_{,\beta}^{(\lambda)}
-u_{\alpha\beta} \right)
\label{eq:Epot2}
\end{array}
\ee
where the Lagrange multiplier $\chi(x)$ is a scalar field.
Taking the functional derivative 
$\delta E/\delta u_{\alpha\beta} =0$ gives 
$\sigma_{\alpha\beta}=(d.c.)_{\alpha\beta}\,\chi$.
Thus $\partial_{\alpha}\sigma_{\alpha\beta}=0$
and $\chi$ is the scalar stress function of Airy \cite{Seung,Lob1}.

In terms of the stress function 
$\kappa_{G}[\chi]=[(1-c_{2})/2\mu]\grad^{4}\chi$
and the geometric von Karman equation is
\be
\frac{1}{\mu}(1-c_{2})\grad^{4}\chi=
-(d.c.)_{\alpha\beta}
(f_{,\alpha}^{(\lambda)}f_{,\beta}^{(\lambda)})
\ee
The force von Karman equation is found via 
$\delta E/\delta f^{(\lambda)}=0$ to be
\be
2\kappa (1+c_{0}) \grad^{4}f^{(\lambda)}=
f^{(\lambda)}_{,\alpha\beta}\,(d.c.)_{\alpha\beta}\,\chi
\ee
which reproduces Eq. (\ref{eq:fvK1}).

\section{The ridge}

\subsection{Boundary Conditions}

In this section we discuss the picture of a ridge
as a {\it boundary layer} and find a simple 
boundary condition which yields a ridge for general $M$. 
Previous analytic studies of
stress confinement have been limited to the case of a 
thin plate ($M =2$) in $\Re^{3}$. 
In Ref. \cite{Lob1}, Lobkovsky treated the case of 
a single ridge in isolation
and analyzed the resulting von Karman equations  
to lowest order in the thickness. He found that 
a ridge of length $X$ has an elastic energy 
$E \sim \mu h^{2} (X/h)^{1/3}$
and a width $w \sim h^{1/3}X^{2/3}$.

We develop the concept of the ridge as a boundary layer
by treating the thickness $h$ as a tunable parameter.
We start with $h=0$. Since the width of a ridge 
scales like $h^{1/3}$, the zero thickness limit
of a ridge is a straight line of zero width. The
geometry of the ridge, shown in Fig. 1,
is two flat plates which meet
at a nonzero angle $D$. The curvatures are obviously 
singular on this line and zero elsewhere. The intuitive
reason for this behavior is that $\kappa/\mu \sim h^{2}$, so
there is no energy cost for curvatures when $h=0$.
When $h$ is made nonzero, the plate
can achieve a lower total energy by smoothing out the singularity
and trading stretching energy
for bending. The resulting balance generates a new length
scale, which is the width of the ridge. This picture will
remain essentially unchanged when $M >2$. The important
point is that the ridge is a boundary layer which
regularizes the $h=0$ singularity.

Lobkovsky began with the semi-infinite strip   
$\{ -\infty < x_{1} <\infty ,
-X/2 \leq x_{2} \leq X/2 \}$. Then he
assumed the presence of (unspecified) normal 
forces acting at the boundary
$x_{2} = \pm X/2$ sufficient to deform the strip
into the ridge shown in Fig. 1.
The boundary conditions are 
$C_{\alpha\beta}(x_{1},\pm X/2)=0$,
$\sigma_{\alpha\beta}(x_{1},\pm X/2)=0$, and
$f(x_{1},\pm X/2)=\alpha |x_{1}|$.
The first two conditions are chosen for convenience.
It is the third condition which determines the shape of
the ridge. The potential $f$ plays the role of the normal
coordinate. The dihedral angle of the resulting ridge is 
$D=\pi -2\alpha$.

The generalization of this geometry to $M >2$ 
is straightforward. We limit our discussion to 
the hypersurface $N = M +1$, since this captures 
the most important features of the general case \cite{Kram3}.
We take for our $h=0$ 
ridge the singular boundary between two M-dimensional
regions with zero curvature. 
As discussed in Ref. \cite{Kram1}, such a boundary 
can have no curvature in the material coordinates
or in $\Re^{N}$. Our boundary conditions must be 
consistent with this ridge.

We take for our undeformed manifold the 
semi-infinite domain 
$\{ -\infty < x_{1} <\infty$ , 
$x_{\bar{\alpha}}\hat{e}_{\bar{\alpha}} \in {\cal C}$
for $\bar{\alpha} \in [2,M] \}$ (see Fig. 2). 
The material coordinate
$x_{1}$ is perpendicular to the ridge and 
${\cal C}$ is an arbitrary, simply connected cross-section.
The ridge is imposed by some
(unspecified) normal forces sufficient to 
create the ``kinked'' potential 
$f|_{\partial {\cal M}}= \alpha |x_{1}|$.
For simplicity we take 
$C_{\alpha\beta}|_{\partial {\cal M}}=0$ and
$\sigma_{\alpha\beta}|_{\partial {\cal M}}=0$. The solution 
to the von Karman equations when $h=0$ is
the singular ridge $f(x)=\alpha |x_{1}|$.

\subsection{Scaling analysis}

To begin a scaling analysis of the von Karman equations
we need to decide which fields to work with.
Since the stress function $\chi_{\alpha\beta}$
is a gauge field in M $>2$, it is convenient to work directly
with the stress tensor $\sigma_{\alpha\beta}$.
We use the bending potential $f^{(1)}$ instead 
of the curvature tensor $C^{(1)}_{\alpha\beta}$
because of the obvious advantages of a 
scalar. Since $N = M +1$ we can drop the normal index.
The geometric and force von Karman equations are then
\be
(d.c.)_{\alpha\beta\mu\nu}
(f_{,\mu}f_{,\nu})
& = & \frac{1}{\mu}(d.c.)_{\alpha\beta\mu\nu}
(\sigma_{\mu\nu}-c_{2}\delta_{\mu\nu}\sigma_{\gamma\gamma}) \\
2\kappa (1+c_{0}) \grad^{4}f & = & \sigma_{\alpha\beta}
f_{,\alpha\beta} \label{eq:fvK3} 
\ee

Now convert to the dimensionless quantities
\be
\check{f}=f/X \mbox{  ,  }
\check{\sigma}_{\alpha\beta}=\frac{X^{2}}{2\kappa(1+c_{0})}
\sigma_{\alpha\beta}
\mbox{  and  }
\check{x}_{\alpha}=x_{\alpha}/X
\ee
where $X$ is a length scale characterizing the 
cross-section ${\cal C}$. Including the conservation law,
the dimensionless equations are 
\be
\partial_{\alpha}\check{\sigma}_{\alpha\beta} & = & 0 
\label{eq:con4} \\
(d.c.)_{\alpha\beta\mu\nu}
(\check{f}_{,\mu}\check{f}_{,\nu})
& = & \epsilon^{2}(d.c.)_{\alpha\beta\mu\nu}
(\check{\sigma}_{\mu\nu}-c_{2}\delta_{\mu\nu}
\check{\sigma}_{\gamma\gamma}) 
\label{eq:gvK4} \\
\grad^{4}\check{f} & = & \check{\sigma}_{\alpha\beta}
\check{f}_{,\alpha\beta} 
\label{eq:fvK4} 
\ee  
where $\epsilon^{2}=2\kappa(1+c_{0})/(\mu X^{2}) \sim (h/X)^{2}$.
As for a bent plate in three dimensions,
we expect the ridge solution to be valid when 
$0 < \epsilon \ll 1$. Note that a naive 
count of Eqs. (\ref{eq:con4})-(\ref{eq:fvK4}) 
gives $M(M+1)/2 + N$ constraints on
$M(M+1)/2 + N-M$ field variables. This set is
not overdetermined because the argument of the 
double curl has the local
gauge freedom $()_{\mu\nu}\rightarrow 
()_{\mu\nu}+(1/2)(\xi_{\mu,\nu}+\xi_{\nu,\mu})$
where $\vec{\xi}(x)$ is an arbitrary vector field.
The geometric von Karman equation therefore represents
only $M(M-1)/2$ independent constraints.

The $\epsilon=0$ limit is the $h=0$ limit. As discussed
in the previous section, the solution to this reduced 
problem is 
the singular ridge $\check{f} = \alpha |\check{x}_{1}|$.
We might hope to solve for the $\epsilon >0$ ridge as a  
perturbation about this solution.
However $\epsilon$ multiplies the highest
derivative of $\check{\sigma}_{\alpha\beta}$. It is therefore
a singular perturbation and naive approaches fail.

To find the exponents which characterize the ridge,
we make the following rescalings
\be
\tilde{\sigma}_{11}=\epsilon^{\delta}\check{\sigma}_{11} \mbox{ , }
\tilde{\sigma}_{1\bar{\alpha}}=\epsilon^{s}\check{\sigma}_{1\bar{\alpha}}
\mbox{  ,  }
\tilde{\sigma}_{\bar{\alpha}\bar{\beta}}=
\epsilon^{t}\check{\sigma}_{\bar{\alpha}\bar{\beta}}
\nonumber \\
\tilde{f}=\epsilon^{\beta}\check{f}  \mbox{  ,  }
\tilde{x}_{1}=\epsilon^{\beta}\check{x}_{1}
\mbox{  and  }
\tilde{x}_{\bar{\alpha}}=\epsilon^{0}\check{x}_{\bar{\alpha}}
\ee
where we distinguish the coordinates parallel to the ridge 
with a barred Greek index $\bar{\alpha}\in [2,M]$.
Note that $\tilde{f} \sim \tilde{x}_{1}$ is required
by the boundary condition, so they scale with the 
same power of $\epsilon$.

To solve for the exponents we start with the conservation 
law Eq. (\ref{eq:con4}). Grouping terms of like order in $\epsilon$ gives
\be
\epsilon^{\beta-\delta}\partial_{1}\tilde{\sigma}_{11}+
\epsilon^{-s}\partial_{\bar{\alpha}}\tilde{\sigma}
_{\bar{\alpha}1} & = & 0
\\
\epsilon^{\beta-s}\partial_{1}\tilde{\sigma}_{1\bar{\beta}}+
\epsilon^{-t}\partial_{\bar{\alpha}}\tilde{\sigma}
_{\bar{\alpha}\bar{\beta}} & = & 0
\ee
Assuming none of the rescaled quantities vanish,
this implies $s=\delta-\beta$ and $t=\delta-2\beta$.

The rescaled force von Karman equation is
\be
(\epsilon^{2\beta}\partial^{2}_{1}
+\epsilon^{0}\partial^{2}_{\bar{\alpha}})^{2}\tilde{f}
=\epsilon^{2\beta-\delta} (\tilde{\sigma}_{\alpha\beta}
\tilde{f}_{,\alpha\beta}) 
\ee
If $\beta \geq 0$ this equation is dominated 
by the lowest order terms in $\epsilon^{\beta}$ 
as $\epsilon \rightarrow 0$ and
$0=2\beta -\delta$. If $\beta  < 0$ then
$4\beta =2\beta -\delta$.

The rescaled geometric von Karman equation has 
$M(M+1)/2$ components, one for each
component of $G_{\alpha\beta}$. For the scaling 
analysis they
can be grouped into four classes $G_{11}$, $G_{1\bar{\alpha}}$,
$G_{\bar{\alpha}\bar{\alpha}}$ (no sum), and 
$G_{\bar{\alpha}\bar{\beta}}$. Assuming none of the relevant
terms vanish, we only need to consider one example 
from each class. The $G_{11}$ component is
\be
\epsilon^{-2\beta}(\tilde{f}_{,\bar{\alpha}\bar{\beta}}
\tilde{f}_{,\bar{\alpha}\bar{\beta}}
-\tilde{f}_{,\bar{\alpha}\bar{\alpha}}
\tilde{f}_{,\bar{\beta}\bar{\beta}})
=-\epsilon^{2-\delta} \{
\epsilon^{2\beta} \tilde{\sigma}_{\bar{\alpha}\bar{\beta},
\bar{\alpha}\bar{\beta}}
+\epsilon^{2\beta}c_{4}\tilde{\sigma}_{\bar{\alpha}\bar{\alpha},
\bar{\beta}\bar{\beta}}
+\epsilon^{0}c_{2}(M-2)\tilde{\sigma}_{11,\bar{\beta}\bar{\beta}}
\}
\label{eq:G11}
\ee
where $c_{4}=1-(M-2)c_{2}$. We have made extensive
use of the conservation law $\tilde{\sigma}_{\alpha\beta,\alpha}=0$
and the symmetries of the problem to simplify the expression.
The other components are derived in Appendix B.
A careful analysis shows that all components yield the same constraint 
on the exponents. If this were not the case, the 
equations for the exponents would be overdetermined.
If $\beta \geq 0$ then Eq. (\ref{eq:G11}) gives
$-2\beta = 2 -\delta$. If 
$\beta < 0$ then $-2\beta = 2 -\delta +2\beta$.

The constraints on the exponents are only solvable if
$\beta <0$. The unique solution is 
\be
\beta=-\frac{1}{3} \mbox{   ,   } \delta=+\frac{2}{3}
\ee 
These are identical to the exponents found by Lobkovsky
for the ridge in ($M=2,N=3$) \cite{Lob1}. There are a variety
of geometric intuitions associated with these exponents.
The most important is that the ridge width is characterized
by the transverse curvature via $w \sim 1/C_{11}$. Thus 
\be
w \sim \epsilon^{-\beta}X \sim h^{1/3}X^{2/3}
\ee
We refer the reader to Ref. \cite{Lob1} for a fuller
discussion of the ridge geometry.

It is instructive to consider 
the rescaled ridge energy to leading order in $\epsilon$ 
\be
\begin{array}{rl}
\displaystyle 
E = \int_{{\cal M}} & (X^{M}\epsilon^{-\beta} d\tilde{x}^{M}) 
\\ & \displaystyle  \left\{
\frac{1}{2\mu}\left(  \frac{2\kappa (1+c_{0})}{X^{2}}
\epsilon^{-\delta+2\beta} \right)^{2}
(\tilde{\sigma}_{\bar{\alpha}\bar{\beta}}^{2}
-c_{2}\tilde{\sigma}_{\bar{\alpha}\bar{\alpha}}^{2})
+\kappa \left( \frac{\epsilon^{\beta}}{X}\right)^{2}
(1+c_{0})\tilde{C}_{11}^{2}  \right\}
\end{array}
\ee
Gathering terms and using the geometric von Karman
constraint $\delta-4\beta=2$ gives 
\be
E = \mu h^{M} (c_{s}\epsilon^{-M-5\beta}+c_{b}\epsilon^{-M+2+\beta})
\label{eq:ebal}
\ee
where $c_{s}$ and $c_{b}$ are dimensionless constants
due to the stretching and bending energy respectively.
In this form it is clear how the value of $\beta$,
and hence the width, is generated
via the balance between strains and curvatures.
As the width $w \sim \epsilon^{-\beta}$ is increased, the 
bending energy decreases and the stretching energy 
increases. The ridge chooses the value
$\beta=-1/3$ which minimizes the total energy. Thus 
\be
E \sim \mu h^{M} \epsilon^{-M+5/3} \sim \mu h^{M} (X/h)^{M-5/3}
\ee
Furthermore, Eq. (\ref{eq:ebal}) fixes exactly the  
ratio $c_{b}/c_{s}$.
At the minimum,
\be
\left. \frac{\partial E}{\partial\beta} \right|_{\beta=-1/3}=
\mu h^{M} \ln (\epsilon)
\left. (-5 c_{s}\epsilon^{-M-5\beta}+c_{b}\epsilon^{-M+2+\beta})
\right|_{\beta=-1/3}=0 
\ee 
This reduces to $c_{b}/c_{s}=5$, which 
means the bending energy is exactly 
five times the stretching energy in an asymptotic ridge.
As noted in the introduction,
this ratio also holds for ($M=2,N=3$) \cite{LobWit}.

\section{Conclusions}

In this paper we derive the equations of static
equilibrium for an $M$-dimen\-sion\-al elastic manifold 
embedded in $N$-dimensional space. We define 
the potentials $\chi_{\alpha\beta}$
and $f^{(\lambda)}$ on the manifold. 
These are the higher-dimensional
analogs of the stress function $\chi$ and 
the bending potential $f$ of a thin plate in
three dimensions. We find a novel interpretation 
for the stress function as the Lagrange multiplier
of the geometric von Karman equation in the elastic
energy functional of the manifold. 

We go on to consider the properties
of an $M-1$ dimensional ridge in an $M > 2$ dimensional
manifold. The scaling is essentially identical to
that found by Lobkovsky for a ridge in $M=2$ \cite{Lob1}. 
We find that a ridge of linear size $X$
in a manifold of thickness $h$ has a width
$w \sim h^{1/3}X^{2/3}$ and a total elastic energy
$E \sim \mu h^{M} (X/h)^{M-5/3}$, where
$\mu$ is a stretching modulus.
The scaling analysis also fixes exactly the ratio
of bending energy to the stretching energy in a ridge
$E_{bend}/E_{stretch} =5$. These results are valid
in the thin limit $h \ll X$. Although our calculations
are explicitly for a bent hypersurface $N=M+1$,
unpublished theory and simulations lead us to expect 
no change in the ridge exponents when $N > M+1$.

The purpose of this work was 
primarily as an aid to future studies of 
crumpling in high dimensional systems.
In particular, in future papers we will discuss the 
phenomenon of spontaneous ridge formation
as a mechanism of stress {\it confinement} \cite{Kram1,Kram3}.
The elastic energy and ridge exponents 
derived here are an essential foundation for that work.
We should point out that the scaling analysis 
in this paper is in no way a guarantee that these 
ridges will form in a crumpled manifold.
The question of how the elastic energy is distributed
is best resolved in combination with computer
simulations and, for ($M=2,N=3$), experiments.
As an example of the way ridge formation can fail,
we note that when $N >2M$ a manifold with a free
boundary can make its stretching energy zero everywhere. 
Since the ridge structure depends on the competition
between bending and stretching energy, no ridge
formation is possible.

\acknowledgments
The author would like to thank
Bob Geroch and Alex Lobkovsky for helpful discussions. 
This work was supported in part
by the NSF through Grants No. DMR-9400379 and DMR-9528957.
This paper was completed in partial fulfillment of 
the requirements for a Ph. D. in physics at the 
University of Chicago under the supervision of
Thomas A. Witten. 

\appendix

\section{Comments on the Taylor expansion}

Begin by expanding $\vec{a}^{(\mu)}$ and $\vec{b}^{(\mu)(\nu)}$
in the full basis $\{ \vec{t}_{\alpha},\hat{n}_{(\alpha)} \}$
\be
\vec{a}^{(\mu)} & = & a^{(\mu)}_{\alpha}\,\vec{t}_{\alpha}
+a^{(\mu)(\nu)}\,\hat{n}^{(\nu)}
\nonumber \\
\vec{b}^{(\mu)(\nu)} & = & b^{(\mu)(\nu)}_{\alpha}\,\vec{t}_{\alpha}
+b^{(\mu)(\nu)(\lambda)}\,\hat{n}^{(\lambda)}
\ee
The coefficients $(a^{(\mu)}_{\alpha},a^{(\mu)(\nu)}, 
b^{(\mu)(\nu)}_{\alpha},b^{(\mu)(\nu)(\lambda)})$ 
are functions of $u^{c}_{\alpha\beta}$, $C^{(\mu)}_{\alpha\beta}$,
$\tau^{(\mu)(\nu)}_{\alpha}$, and their derivatives.
We require that the expressions for the coefficients
in terms of these quantities have the correct
number of free manifold and normal indices
(there may be an arbitrary number of contracted
indices). This is necessary and sufficient for their
correct behavior under reflections and rotations
of the manifold and normal coordinates.

Note that $b^{(\mu)(\nu)}_{\alpha}$ and
$b^{(\mu)(\nu)(\lambda)}$ must be even under
$\mu\leftrightarrow\nu$ since 
$\vec{b}^{(\mu)(\nu)}(x)=\partial_{(\mu)}\partial_{(\nu)}
\vec{r}(x,\zeta)|_{\zeta=0}$. This is why
$b_{3}\tau^{(\mu)(\nu)}_{\alpha}\,\vec{t}_{\alpha}$
is not a valid term. Note also
that $\tau^{(\mu)(\mu)}_{\alpha}=0$.

We require that the coefficients have the correct 
units. Consider the following additions to the expression for 
$\vec{a}^{(\mu)}$
\be
\vec{a}^{\prime (\mu)}=(\tilde{a}_{2}C^{(\mu)}_{\alpha\beta}
C^{(\nu)}_{\alpha\beta} + \tilde{a}_{3}
\tau^{(\mu)(\lambda)}_{\alpha}\tau^{(\lambda)(\nu)}_{\alpha})
\,\hat{n}^{(\nu)} + (\tilde{a}_{4}\tau^{(\mu)(\nu)}_{\alpha}
C^{(\nu)}_{\alpha\beta}+\tilde{a}_{5}C^{(\mu)}_{\alpha\beta,\beta})
\,\vec{t}_{\beta}
\ee
Since $\vec{a}^{(\mu)}$ is dimensionless
the constants $\tilde{a}_{j}$ must have units
of $(\mbox{length})^{2}$. The only length scale 
available is the thickness $h$, so we 
write $\tilde{a}_{j}=a_{j}h^{2}$. The expressions
are therefore of $O(h^{2}C^{2},h^{2}\tau^{2},h^{2}C\tau,
\mbox{  and  } h^{2}C/\ell)$ respectively. 
Although these terms are assumed small,
they are not necessarily negligible compared 
to $u^{c}_{\alpha\beta}$. One can verify, however,
that their contribution to the energy is negligible
compared to the curvature and torsion
terms in Eq. (\ref{eq:Etot}). Similar 
arguments lead to the form for $\vec{b}^{(\mu)(\nu)}$.

In the theory of thin shells it is known that the centroid
deformations alone are not sufficient to describe the
behavior near the boundary of the shell.
The full three-dimensional problem must be solved there.
As a consequence, any energy functional derived
via a Taylor series expansion in the thickness is not
uniformly convergent near the boundary.
For a detailed discussion of these considerations,
we refer the reader to Ref. \cite{Shell2} and the
references therein.

\section{Scaling of the force von Karman equation}

Assuming none of the relevant terms vanish, we only
need to consider one example from each of the three
classes $G_{1 \bar{\alpha}}$, $G_{\bar{\alpha}\bar{\alpha}}$
(no sum), and $G_{\bar{\alpha}\bar{\beta}}$.
Dropping the tilde notation, the $G_{12}$ equation is 
\be
\begin{array}{rcl}
2 \epsilon^{-\beta}(f_{,12}f_{,\bar{\alpha}\bar{\alpha}} 
& - &  f_{,1\bar{\alpha}}f_{,2\bar{\alpha}})
\\
& = & -\epsilon^{2-\delta}(\epsilon^{3\beta}\sigma_{12,11}
+\epsilon^{\beta}\sigma_{12,\bar{\alpha}\bar{\alpha}}
+\epsilon^{\beta}c_{4}\sigma_{11,12}
+\epsilon^{3\beta}c_{4}\sigma_{\bar{\alpha}\bar{\alpha},12})
\end{array}
\ee
Defining dotted Greek indices 
$\dot{\alpha}\in [3,M]$, the $G_{22}$ equation is 
\be
\begin{array}{rcl}
2 \epsilon^{0}  (f_{,1\dot{\beta}}f_{,1\dot{\beta}}
& - & f_{,11}f_{,\dot{\beta}\dot{\beta}}) 
 +\epsilon^{-2\beta} (f_{,\dot{\alpha}\dot{\beta}}
f_{,\dot{\alpha}\dot{\beta}} 
-f_{,\dot{\alpha}\dot{\alpha}}
f_{,\dot{\beta}\dot{\beta}})
\\
& = & -\epsilon^{2-\delta} \{
c_{4}(\epsilon^{2\beta}\sigma_{11,11}
+\epsilon^{0}\sigma_{11,\dot{\beta}\dot{\beta}}
+\epsilon^{4\beta}\sigma_{\bar{\alpha}\bar{\alpha},11}
 +\epsilon^{2\beta}\sigma_{\bar{\alpha}\bar{\alpha},
\dot{\beta}\dot{\beta}}) 
\\
&& + \epsilon^{4\beta}\sigma_{22,11}
+\epsilon^{2\beta}\sigma_{22,\bar{\alpha}\bar{\alpha}}
\}
\end{array}
\ee
and the $G_{23}$ equation is 
\be
\begin{array}{rcl}
2 \epsilon^{0} (f_{,23}f_{,11} & -& f_{,21}f_{,31})
+2 \epsilon^{-2\beta} (f_{,23}f_{,\dot{\beta}\dot{\beta}} 
-f_{,2\dot{\beta}}f_{,3\dot{\beta}})
\\
& = & -\epsilon^{2-\delta}(\epsilon^{0}c_{4}\sigma_{11,23}
+\epsilon^{2\beta}c_{4}\sigma_{\bar{\alpha}\bar{\alpha},23}
+\epsilon^{4\beta}\sigma_{23,11}
+\epsilon^{2\beta}\sigma_{23,\bar{\alpha}\bar{\alpha}})
\end{array}
\ee
If $\beta \geq 0$, then all three equations give
the constraint $-2\beta =2-\delta$ in agreement
with the $G_{11}$ component. If $\beta < 0$, then
all three give $0=2-\delta+4\beta$, again in agreement
with $G_{11}$.

\begin{figure}
\centerline{\epsfxsize=8.5cm \epsfbox{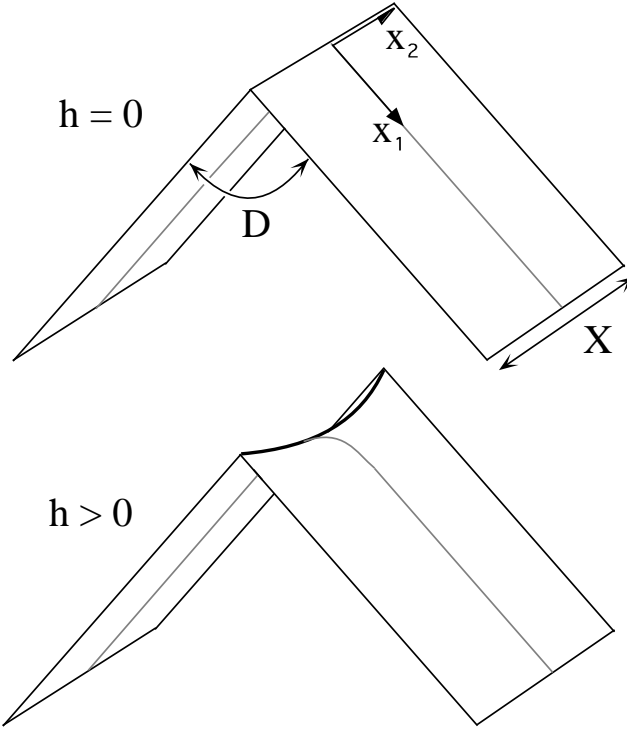}}
\caption{The ridge in a semi-infinite strip
when $h=0$ and $h>0$. We have labeled the 
$h=0$ ridge to show the ridge length $X$,
the dihedral angle $D$, and the manifold coordinate 
system $(x_{1},x_{2})$.}
\label{fig1}
\end{figure}

\begin{figure}
\centerline{\epsfxsize=11.0cm \epsfbox{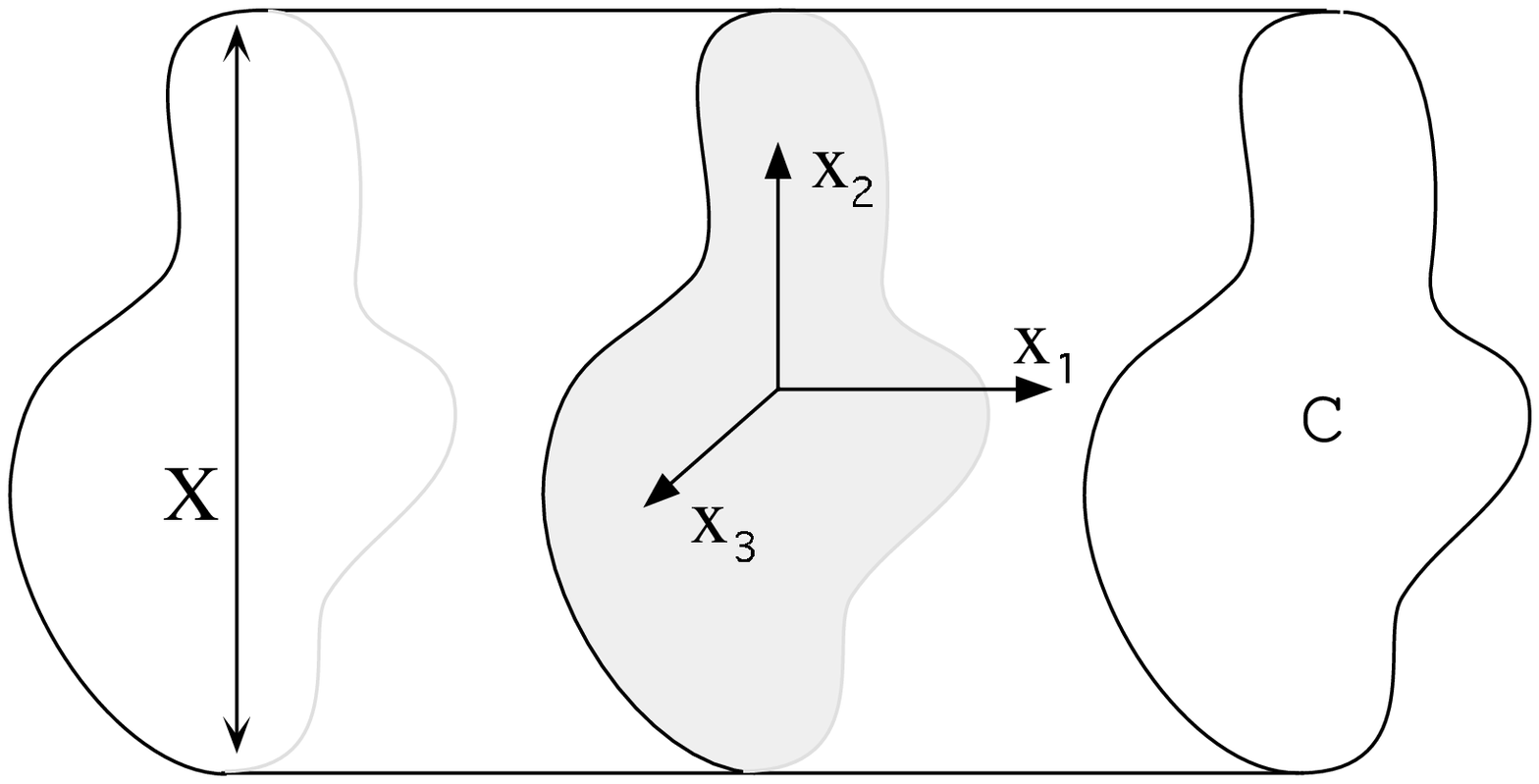}}
\caption{The manifold coordinate system 
$(x_{1},x_{2},x_{3})$ for the semi-infinite
rod $\Re^{1}\times{\cal C}$. The location of the 
planar ridge is indicated in grey.}
\label{fig2}
\end{figure} 


\clearpage 
\begin{references}

\bibitem[\dag]{byline} e-mail address: kramer@rainbow.uchicago.edu
\bibitem{Kram3} E. M. Kramer and T. A. Witten, 
to be submitted to Phys. Rev. Lett.
\bibitem{Crmpl} F. Plouraboue and S. Roux, Physica A (1996).
\bibitem{Car} R. E. Newton, in {\it Seventh International 
Technical Conference on Experimental Safety Vehicles} 
(US Dept. of Transportation, Washington, DC, 1980).
\bibitem{Sound} E. M. Kramer and A. E. Lobkovsky, Phys. Rev. E
{\bf 53}, 1465 (1995).
\bibitem{WitLi} T. A. Witten and Hao Li, Europhysics Letters
{\bf 23}, 51 (1996).
\bibitem{Lob1} A. E. Lobkovsky, Phys. Rev. E {\bf 53}, 3750 (1996).
\bibitem{Lob2} A. E. Lobkovsky {\it et al.}, Science {\bf 270}, 
1482 (1995).
\bibitem{Corr} Surprisingly, the leading
corrections to the ridge are not confined.
See Ref. \cite{LobWit}.
\bibitem{Kram1} E. M. Kramer and T. A. Witten, 
to be submitted to Phys. Rev. E.
\bibitem{MemInt} {\it Statistical Mechanics of Membranes and Surfaces},
edited by D. Nelson, T. Piran, and S. Weinberg (World
Scientific, Singapore, 1989).
\bibitem{CT1} M. Paczuski, M. Kardar, and D. Nelson,
Phys. Rev. Lett. {\bf 60}, 2638 (1988).
\bibitem{CT2} E. Guitter {\it et al.}, J. Phys. France {\bf 50},
1787 (1989).
\bibitem{Love} A. E. H. Love, {\it A Treatise on the Mathematical 
Theory of Elasticity} (Dover, New York, 1944).
\bibitem{LL} L. Landau and E. Lifshitz, {\it Theory of Elasticity}
(Pergamon Press, New York, 1959).
\bibitem{Shell1} {\it The Theory of Thin Elastic Shells}, edited
by W. T. Koiter (North-Holland Publishing, New York, 1960).
\bibitem{Shell2} M. Dikmen, {\it Theory of Thin Elastic Shells}
(Pitman Advanced Publishing, Boston, 1982). See in particular
pp. 156-7 and references
therein for a discussion of convergence questions.
\bibitem{Shell3} J. E. Lagnese, {\it Boundary Stabilization of Thin Plates}
(SIAM, Philadelphia, 1989).
\bibitem{vK} T. von Karman, {\it The Collected Works of 
Theodore von Karman} (Butterworths Scientific Publications, London,
1956), Vol. 1, p. 176.
\bibitem{Airy} G. B. Airy, Phil. Trans. Roy. Soc. {\bf 153}, 49 
(1863). See also the historical introduction in Ref. \cite{Love}. 
\bibitem{Seung} H. S. Seung and D. R. Nelson, Phys. Rev. A
{\bf 38}, 1005 (1988).
\bibitem{Maxwell} J. C. Maxwell, Edinburgh Roy. Soc. Trans.
{\bf 26} (1870), reprinted in {\em The Scientific 
Papers of James Clerk Maxwell}
(Cambridge University Press, Cambridge, 1927), Vol. 2, p. 161.
\bibitem{Kleinert} H. Kleinert, {\it Gauge Fields in Condensed Matter}
(World Scientific, New Jersey, 1989).
\bibitem{LobWit} A. E. Lobkovsky and T. A. Witten, to be submitted
to Phys. Rev. E.
\bibitem{Diff1} L. P. Eisenhart, {\it Differential Geometry}
(Ginn and Company, New York, 1909).
\bibitem{Diff2} R. S. Millman and G. D. Parker, {\it Elements 
of Differential Geometry} (Prentice-Hall, New Jersey, 1977).
\bibitem{GR} B. Schutz, {\it A First Course in General Relativity}
(Cambridge University Press, Cambridge, 1988).
\bibitem{Stegun} M. Abromowitz and I. Stegun, {\it Handbook 
of Mathematical Functions} (Dover, New York, 1972).
\bibitem{Monge} See sections 11-14 of Ref. \cite{LL} for a complete
treatment of a thin plate in the Monge representation.
\bibitem{gauge} Note that the energy functional is invariant
under this transformation. One integration by parts
moves the derivative onto $G_{\alpha\beta}$.
\end{references}
\end{document}